\documentclass[12pt]{article}
\usepackage{graphicx}
\usepackage{epsfig}
\usepackage{latexsym}
\usepackage{amsmath}
\usepackage{xcolor}

\begin{document}

\title{   
Benford's law first significant digit and distribution distances for  testing the  reliability of  financial reports in  developing countries 
  }

\author{Jing Shi\footnote{$e$-$mail$ $address$: 248299825@qq.com}   
 \; and Marcel Ausloos\footnote{ $e$-$mail$ $address$:
ma683@le.ac.uk;  also at 
GRAPES,  483 rue de la belle jardini\`ere, B-4031 Li\`ege, Euroland; $e$-$mail$ $address$:
marcel.ausloos@ulg.ac.be;  
} \\   School of Business, Ken Edwards Building, \\University of Leicester, Leicester, LE1 7RH, United Kingdom 
\\  \\ Tingting Zhu\footnote{$e$-$mail$ $address$:  tingting.zhu@dmu.ac.uk}     
 \\Faculty of Business and Law,  Hugh Aston Building, \\De Montfort University,  Leicester, LE1 9BH, United Kingdom }

 \date{\today}
\maketitle \begin{abstract}
We discuss 
a common suspicion about  reported financial data, in 10 industrial sectors of the 6   so called "main  developing countries" over the time interval [2000-2014]. These data are examined through  Benford's law first significant digit and  through distribution distances tests.

 It is shown that several  visually anomalous data  have to  be {\it a priori} removed. Thereafter,  the distributions much better follow the first digit significant law, indicating the usefulness of a Benford's law test from the research starting line.  The same holds true for distance tests. A few outliers are pointed out.
 \end{abstract}


\section{Introduction  }\label{sec:intro}

 The Benford's law is a phenomenological law about the probability distribution of  the first significant digits (henceforth FSD) in a data set \cite{Nigrini2012}.  
  Newcomb  (in 1881) and later Benford (in 1938)  \cite{ref[1],ref[2]} observed  that the occurrence of the first significant digit  in many data sets is not
uniform, but tends to follow a logarithmic distribution such that the smaller digits appear as  a  first significant digit  more frequently than the larger ones,   according to  
\begin{equation}\label{Beneq1}
N_{d}= N\; log_{10}(1+\frac{1}{d}), \;\;\; d \equiv 1, 2, 3, . . . , 9
\end{equation}
where $N$ is the total number of considered 1-st digits, in short, the number of data points,  and  $N_{d}$ is the number of the so observed integer $d$ ( $= 1, 2, 3, . . . , 9$)  being the starting one (1-st) in the data set list \cite{PS8.11.1}.

 The law,  Eq. (\ref{Beneq1}), nowadays called Benford's law  (BL),  is  widely applied in the investigation of data manipulation by researchers in  finance and economy  
 \cite{Mebane08,JudgeSchechter,EPJB87.14.261MirRCMABenfordIT}.
The BL  can be used not only to identify falsely created data, e.g. in corporationsÕ financial statements   as shown by Nigrini \cite{Nigrinibook},  
 but also  to  verify  the (non)reliability of macroeconomic data 
  \cite{GER12.11.243}.    Furthermore, because of its special features, BL has been employed as a quality  test criterion for  various employed data. 
 An extensive bibliography, from 1881 up to 2006,  on Benford's law papers including theories, applications, generalizations and warnings can be found in a H\"urlimann 
   unpublished work but available on internet  \cite{0607168befordlawbiblio}; see also    books by Kossovsky 
    for more  recent reviews  \cite{Kossovsky14,Kossovsky17}.

In particular,   the Benford's law test  (henceforth BL test)  has been   applied to many financial numerical series. Concentrating on stock market indices, let us mention  (in a chronological order)  Ley (in 1996)  \cite{Ley96} with a 
 Bayesian approach  observed  the   distribution of  U.S. stock indexes' digits:  the series of one-day returns,  for  the Dow Jones Industrial Average (DJIA) and the Standard and Poor's Index (S\&P),   were found to follow Benford's Law. Ley  interestingly concluded that distributions that follow Benford's Law are distributions where small changes are more likely to occur than large changes.
 The  FTSE\footnote{Financial Times Security Exchange} 100  was examined by  De Ceuster et al. (in 1998)  \cite{DeCeusteretal98}. 

After about a 10 years lapse with no investigation of the sort,  
Krakar  and  \v{Z}gela (in 2009)  studied the Zagreb Stock Exchange and found out that the closing prices did not follow BL \cite{KrakarZgela09}.  

Corazza, et al. (in 2010) checked the S\&P 500    \cite{Corazzaetal10}, while 
 Zhao and Wu (also in 2010) wondered whether   the Shanghai Stock Exchange  Composite index and  the  Shenzhen Stock Exchange Component   index  agree with Benford's Law, finding that  Benford's law reasonably holds for these two (main) Chinese stock indices  \cite{ZhaoWu10}.
 
 \v{Z}gela (in 2011) analysis of DAX\footnote{Deutscher Aktien IndeX} percentage changes   over 10 years  [2001-2011] led to the conclusion that DAX values were not in
accordance with the  FSD Law \cite{Zgela11}.
More recently,  the   Istanbul Stock Exchange (BIST) attracted  some attention:  Karavardar (in 2014) \cite{Karavardar14}    found  no disagreement with BL  for the monthly returns  (over 26 years),   while  Cinko (2014)   found no disagreement for the daily returns   (over 23 years) \cite{Cinko14}.
 
 Under the efficient market hypothesis, the stock price index should be completely controlled by the market, thus  would not be manipulated by human (or government) intervention. Under this situation, we could use the BL test to detect if faults exist in the price indexes 
 
For completeness, let us mention   opinions wondering whether BL has to be ever applied, e.g.  as discussed by Mebane   \cite{Mebane08}  or Durtschi et al.    \cite{Durtschietal04}.  An argument on the non-reliability of detecting fraud by BL is based on the knowledge of BL by  potential swindlers, aware of the necessity of conforming to the BL.  In fact, such a possibility might not be surprising,  or is even  sustained,  in view of findings on income taxes regularity in some Italian provinces  by Mir et al.
\cite{EPJB87.14.261MirRCMABenfordIT,CSF104.17.238MARCTAM/ATI_BL}. 
  Such an argument suggests  to consider  alternative methods complementing the  BL classical analysis 
   \cite{NigriniMiller09,ClippeAusloosTheil,Gauvrit017AdvCognPsy}.
 
We propose here a data aggregation method (see also Mir et al. \cite{EPJB87.14.261MirRCMABenfordIT})  which could prove  to be useful, especially in view of the fact that it would be particularly difficult to fabricate data conforming to  BL by various unconnected agents along this scheme.

We have conducted the BL  test  in order  to evaluate whether the collected Financial Times Security Exchange (henceforth FTSE)  global price index of some financial data possibly contains  error values.   We   have examined the case of  monthly log-returns for six (top, -  from a   Gross Domestic Product (GDP) point of view)   emerging countries, over 10 different industrial sectors, - over 15 recent years,  as reported by FTSE.    
   There is  no study to our knowledge, about data  from emerging (or developing) countries, on the scale endeavored here below, but we mention that the GDP  growth rate was studied  for Germany by Gunnel and T\"odter, in 2009,   \cite{GunnelTodter09}. 

 
 Therefore, the  present paper goes as follows: after quickly outlining the BL test, in Sect. \ref{sec:BL},  we complement   the    methodology introducing a  "complementary distance study" \ref{sec:distance}.
In Sect. \ref{sec:dataset}, after explaining the countries selection (Sect. \ref{countries}),   the raw data acquisition  is recalled, i.e. monthly log-returns, next aggregated over 6 relevant emerging countries for a  15 year    time interval (Sect. \ref{finAnt}). The data of interest is displayed through histograms   in Sect. \ref{Benrawdata}.   

We have performed the  $\chi^2$  tests  in order to compare the BL to  the observed distributions.  The results appear unsatisfactory. However, a  visual review indicates anomalous values in the raw data. We have removed them and redone  $\chi^2$  tests  with respect to  BL test values;  a comparison of results before and after removal of anomalous data is   presented in Sect.  \ref{Benadapteddata}.


A discussion, practically oriented on BL and on the distribution distance to BL  is found  in Sect.  \ref{BLanal}. 
 In addition,  since we also  investigate a complementary quantity, the distance between the  10  industry price indexesÕ FSD distributions  and BL is  discussed  in Sect. \ref{distanceanal}.   Some  synthesis  allows for a discussion  in Sect. \ref{subdiscussion}. Sect. \ref {sec:conclusions} serves for a conclusion. 
 

\section{Methodology}\label{sec:methodology}

\subsection{Benford's law}\label{sec:BL}

Benford's law, Eq.(\ref{Beneq1}),  is  known as the  ÔÔfirst digit lawÕÕ or  the ÔÔlaw of the leading digitsÕÕ. According to Eq.(\ref{Beneq1}), in a given data set
the probability of occurrence of a certain digit as the first (1-st) significant digit decreases logarithmically as the value of the digit
increases from 1 to 9. Thus, digit 1 should appear as the first significant digit about 30.10\% times,  and
similarly 9 should appear about 4.58\% times.

 Benford's law, Eq.(\ref{Beneq1}),  holds for  data sets in which the occurrence of numbers is free from any restriction; 
 significant deviations from the Benford distribution may indicate fraudulent or corrupted data   
  \cite{ref[45]}.   

 Deviation from the expected Benford's law distribution  is calculated from the  $\chi^2$ statistic according to 
 \begin{equation}
 \label{eq:1BLchi2b}
\chi^2_{1BL} =\sum_{i=1}^9 \frac{(e_{i1}-b_{i1B})^2}{b_{i1B}},
\end{equation}
 where $e_{i1}$ is the observed frequency of each FSD in the price index data;  $b_{i1B}$ is the frequency expected from  Benford's law. The 10\%, 5\%, and 1\% critical values for  $\chi^2,$    with 8 degrees of freedom,  are 13.36, 15.51, and 20.09. 
 
 
 In practice, applications of Benford's law for fraud detection routinely may use more than the first digit   
 \cite{Nigrini96}. Indeed,  Eq. (\ref{Beneq1}) can be generalized to forecast how many times any digit, or any combination of digits, should be found at some rank in the number  
  \cite{MirAusJAIST}.  Moreover,  undetectable correlations by means of the  classical covariance-based measure can be identified in the statistics of the corresponding first digits, as recently  shown by Gramm et al.  \cite{GrammetalPRE17}.
We will not go beyond the FSD  in this report, - going beyond seems indeed that we would mean to pursue a deleterious way on the data; this is outside our aim.

 \subsection{Distribution distance}\label{sec:distance}
 
The question on reliability of data along BL can be debated on theoretical grounds as mentioned here above   \cite{Mebane08,JudgeSchechter,Durtschietal04}. 
In order to complement such  a study we propose to examine  another  measure, - the distance between distributions  as discussed by Cho and Gaines or by Miskiewicz \cite{ChoGaines,JM10}. A distance measure can be  constructed as the Euclidean distance between two distributions:   this distance measure $d^*$ is  defined as
       \begin{equation}
       d^*=\frac{1}{M} \sqrt{\sum_{i=1}^9 (e_{i1}-b_{i1B})^2 },
       \end{equation}
      where $M$ is the maximum possible distance
      
 \begin{equation}
    M= Max_{i=1, ...,9}[ |e_{i1}-b_{i1B}| ] .
    \end{equation}
    
   We  also introduce  a distance measure $ a^*$,  as the absolute value of the difference between the mean of the investigated FSD distribution and the mean of BL  FSD distribution divided by the maximum possible difference. 

 It is worth mentioning that high values of  $\chi^2$,  $M$, $d^*$,  and $a^*$ indicate   weak  similarities between the tested distribution and the theoretical BL distribution, while a large correlation coefficient reflects  that the data FSD distribution is  rather  similar to the Benford's Law distribution.
 
 \section{  Data set   }\label{sec:dataset}

\subsection{Country selection}\label{countries}

To perform an interesting selection, several emerging countriesÕ nominal GDP for 2014  were first  looked at.  The  GDP data    for the countries which we examined can  be  taken from  the  International Monetary Fund's World Economic Outlook Database 2015
   \cite{IMF2015}. It is found that the nominal GDP of China, Brazil, India, Russia, Mexico, Indonesia, Turkey and Saudi Arabia are all larger than six hundred millions USD, i.e. those countries nominal GDPs rank higher than   all other developing countries. Therefore,  each of these eight countries economic power is comparatively stronger than that of any other developing countries. Thus, those countriesÕ financial data appear to provide  a   valuable representativeness for  the  whole set  of  developing country economic sectors. However, we could not take Russia and Saudi Arabia   into account due to   some data unavailability; we could not find suitable proxies to reflect industries returns.

\subsection{Financial data}\label{finAnt}

Recall that we  implement a test on the reliability of  (log-return) values obtained through aggregated data from  emerging countries across various industry sectors. 

The benchmark, which might be the most acceptable industry classification approach, divides all traded equities into ten different industry categories. We follow the
 FTSE industry classification: basic materials (MATS), consumer goods (GDS), consumer services (SVS), financials (FIN), health care (HEA), industrials (INDU), oil and gas (OIL), technology (TECH), telecommunications (TELE), and utilities (UTIL). All 10 industries sections  are investigated in this research. 
In order to measure each industry  stock return, we employ the FTSE Global monthly price index ($http://www.ftse.com/products/indices/geis-series$).  We   transfer them into log returns. All the return rates are reported as percentages.

It is  also worth pointing that there are some (known or not)  limitations in the FTSE   index. First,  it only employs some specific portfolios to represent the   whole market.
Second, the  FSTE price index is not available for several countries' specific industry sections. For example, there is no FSTE price index for Mexico's  technology industry.  
 Because of those reasons, we limited the selection to China, Brazil, India, Mexico, Indonesia and Turkey. The  number of data points $N_s$ depends on the sector $s$.
 
 We also chose   monthly data as our relevant set for the following reasons: on one hand, monthly data is most frequently used in  (related) previous studies; on the other hand, monthly data is more available than daily or quarterly data because  numerous  statistics departments employ the month as  the time interval in their comparative measures,   are easier to  manipulate in view of identifying changes in trends, and better for strategic long term forecasting.  
 
 All data used in this research was downloaded from Datastream. For example, we collected data of FTSE China basic material (MATS) price index, and computed the monthly log return   to represent the return of China's  basic material industry. We did so for the 6 countries and the various industry sectors for the January 2010 - December 2014 time span.  For uniformity, the currency unit of the collected price indexes was  turned into U.S. dollar. In so doing we have a coherently aggregated  data set on which BL tests can be performed.

   \subsection{Histograms of raw data}\label{Benrawdata}
 
\begin{figure}
\begin{centering}
\includegraphics [height=18.5cm,width=22.0cm]
{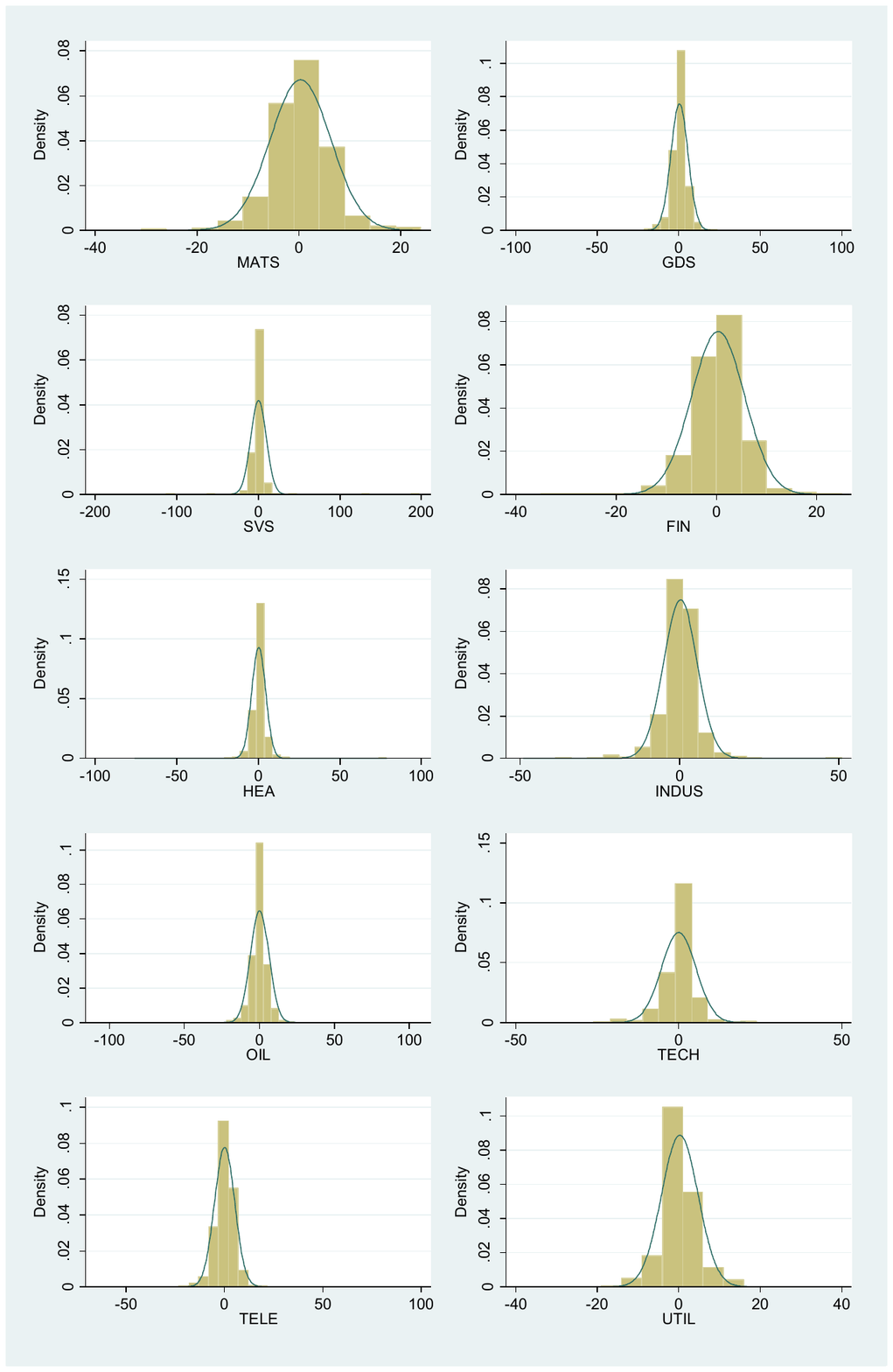}
\caption   {
Distributions  as histograms of the 10  industriesÕ monthly log return over  6  developing countries between January 2000 and December 2014.  } 
 \label{fig:Fig2JingShi}   \end{centering}
\end{figure}

 After  collecting the data for the six countries monthly log returns in the 10  sectors for the time span[Jan. 2000 - Dec. 2014], one can draw the histograms of the dataset.
  Figure \ref{fig:Fig2JingShi}  displays the histograms of the  10 industries monthly log return  aggregated over countries and the considered time interval. The descriptive statistics of data  is found in Table \ref{T4}.

  Note that  the raw data extends over  several orders of magnitudes. 
  For example, the spreads [Max - min]  extend between $\sim 55$  (for UTIL) to $\sim 300$ (for SVS).  Industries, such as consumer services (SVS) and oil and gas (OIL), have maximum returns larger than 100 percent. Moreover, the minimum return of consumer services (SVS) is -105.8 percent, -  which is smaller than -100 percent.  The value range of industry returnsÕ standard deviations is  wide: $\sim 6$, and even $\sim 9.5$ for SVS. Notice that  it practically means that the return of this industry is less stable than that of others within the research period.  Those phenomena are uncommon for stock returns. 
Since there are several abnormal statistical features  the histograms of those industry returns have relatively high peaks and thin tails. The coefficient of variation (StdDev/mean)  extends between $\sim 11$ for GDS up to about $\sim 300$ for TECH.


    \subsection{  Histograms of adapted  data}\label{Benadapteddata}
    
\begin{figure}
\includegraphics [height=18.5cm,width=22.0cm]
{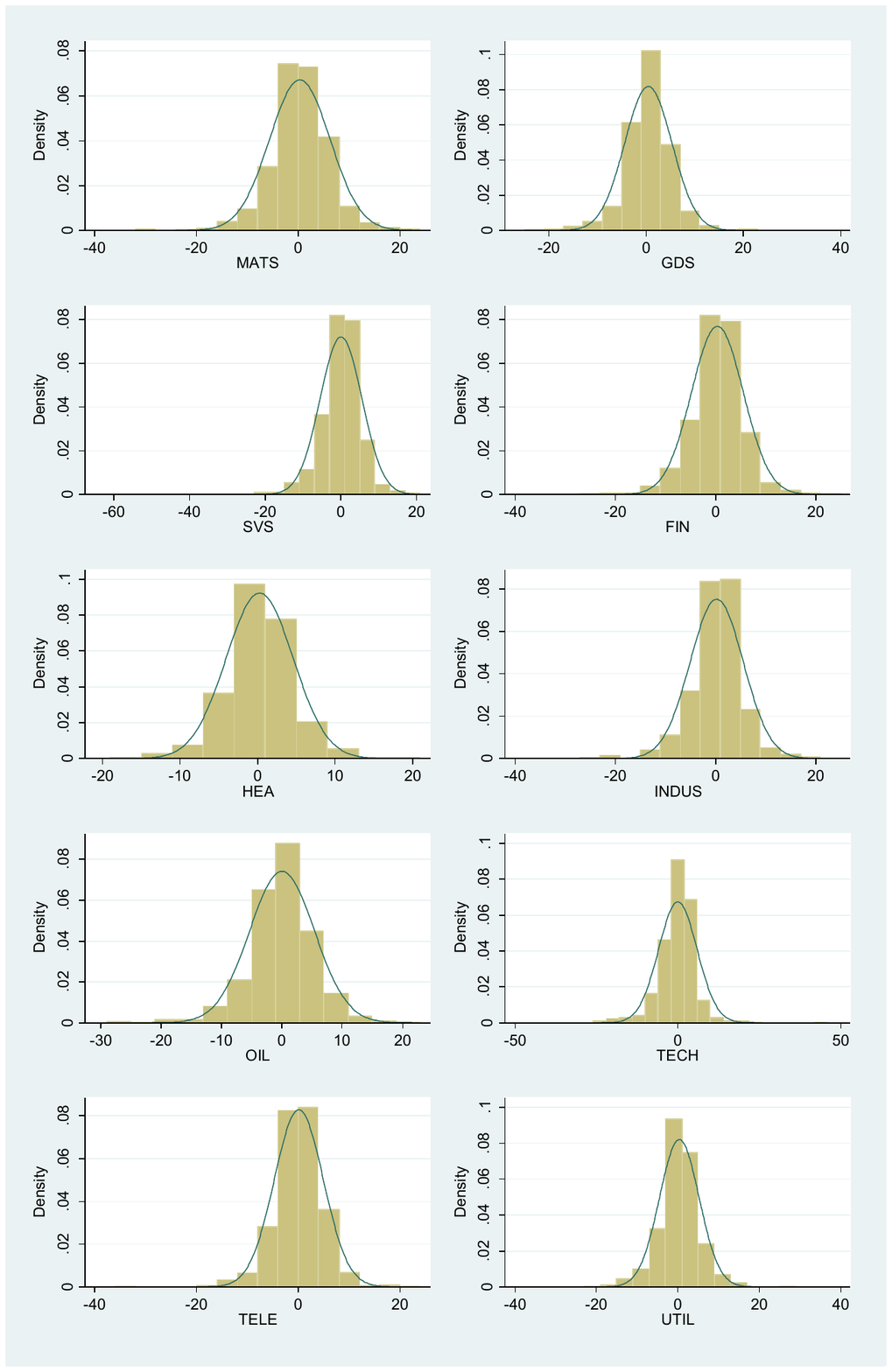}
\caption   {
Histograms of the 10  industries' monthly log return distributions ,  after deletion of abnormal repetitious numbers, seen to be better fitted by a Gaussian} 
 \label{fig:Fig3JingShi} 
\end{figure}

As  anticipated,     abnormal repetitious values exist in the raw data.  An example, - extracted from  the whole data set, is provided in Table \ref{T10}, 
  covering relevant information on China 10 sectors between June 2000 and May 2001. This  example visually demonstrates  abnormal repetitions  in the data set. They are  emphasized  with  bold font. The table blanks  also show some  unavailable data. 
  No certified interpretation of such repetitions could be found.  We suspect that those repetitious numbers are  error values  due to technical problems of Datastream. Consequently, we checked the daily data of the same price indexes in the same period to justify our speculation. We observed  that  repetitions still exist in the same time periods.   Hence, those   values which would generate a biased result of the empirical test have been  deleted.   The descriptive statistics of such "adapted data" is  found in Table \ref{T5}.   Figure \ref{fig:Fig3JingShi}  displays the corresponding  histograms of the  adapted  monthly log return for the 10 industries,   aggregated over the 6 countries, and for  the considered   15 years  time interval.

   After such a data manipulation,  the distribution characteristics are unchanged for MATS and  quasi unchanged for TECH. However, the spreads [Max - min]  extend  now between $\sim 36$  for HEA to $\sim 82$ for SVS.  Industries, the maximum returns are not larger than 45 percent (for TECH). Moreover, the minimum return of consumer services (SVS) is -62 \% now.  The value range of industry returnsÕ standard deviations is  narrowed:  as "low" as $\sim 4.3 $ for HEA, and as  "high" as $\sim 6$ for TECH and MATS.  The coefficient of variation   extends between $\sim -230$ (for SVS, for which the mean takes a negative value),  to about $\sim 245$ (for TECH).

   In order to  emphasize our   data manipulation   rationality, we have conducted the BL and distance tests on the   adjusted data, but also on the raw FTSE  aggregated  price indexes. 

\section{Benford law test and distance test}\label{sec:discussionBL}

Thereafter, the  data can be analyzed along BL tests  through a $\chi^2$  as follows. 

\begin{figure}
\centering
\includegraphics [height=17.5cm,width=22.0cm]
{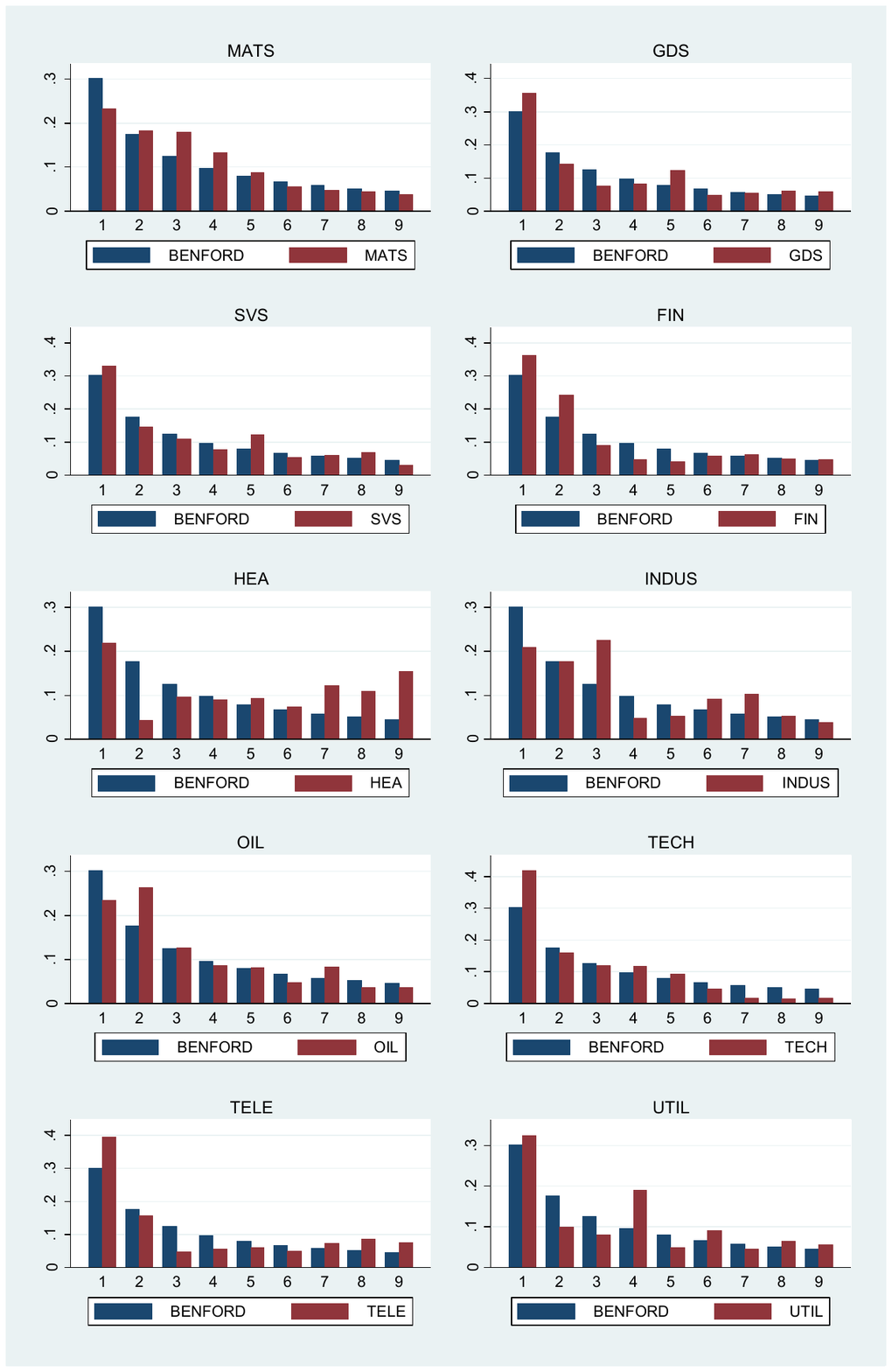}
\caption   {
Comparison through bar charts of 10 industry price indexesÕ FSD frequencies  (red / light) and the expected frequency according to Benford's law (dark /  blue) } 
 \label{fig:Fig4JingShi} 
\end{figure}

 
 Figure \ref{fig:Fig4JingShi}   illustrates the  10 industry price indexesÕ FSD frequencies and the expected frequency according to BL. We drew those bar charts with the frequencies reported in  Table \ref{T11}. The   somber (red) bars represent the observed distributions, the   dark (blue)  bars represent the Benford's law theoretical distribution. The bar charts  show  that all industriesÕ FSD distribution  are  different from the expected BL FSD distribution.

\begin{figure}
\centering
\includegraphics [height=17.5cm,width=22.0cm]
{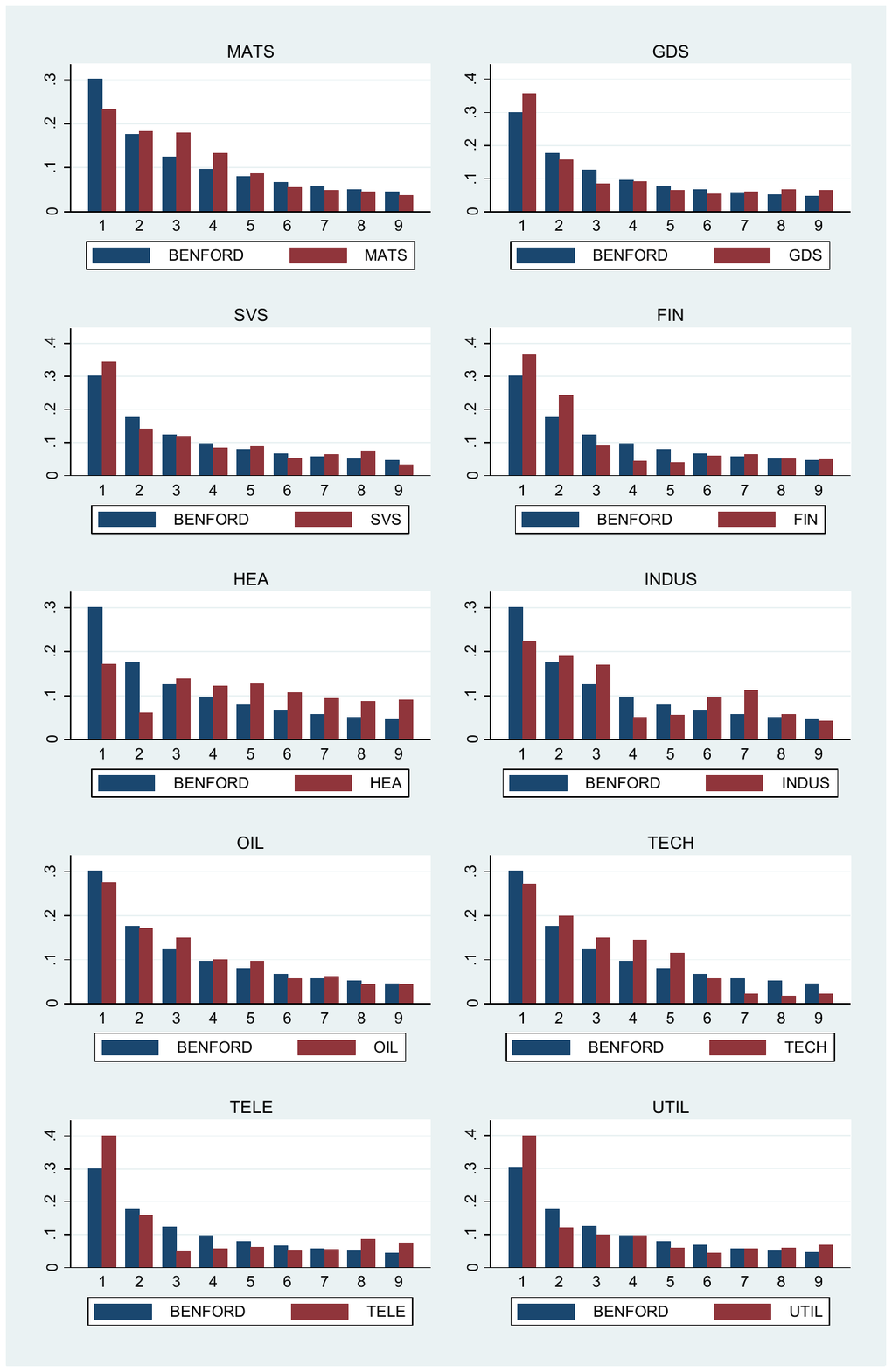}
\caption   {
Comparison by bar charts of  the 10 industry price indexesÕ FSD frequencies  (red / light) and the expected frequency according to Benford's law  (dark /  blue)after the deletion of abnormal repetitious numbers.} 
 \label{fig:Fig5JingShi} 
\end{figure}

 
Beside that, Figure \ref{fig:Fig5JingShi}   illustrates  the same type of data, 
 but after the deletion of the abnormal repetitious numbers;  the frequencies are reported in Table \ref{T13}. The somber (red) bars  again represent  the observed distributions, the dark (blue) bars represent  the BL  expectation. The bar charts shows that all industriesÕ FSD distribution are still different from BLFSD distribution. However, comparing to the distributions before adjustment, those distributions are more similar to the expected BL distribution.
 
  \subsection{ Goodness of fit test   analysis} \label{BLanal} 
 
How much the distributions of the first digits  match the distribution specified by  BL, Eq.(\ref{eq:1BLchi2b}),  can be now tested through a $\chi^2$ test.  
  The results for the raw data and for  the adapted data are found in Table \ref{T12} and Table \ref{T14}, respectively. The  $^*$$^*$  indicates a 99\% significant departure from BL. 


Comparing Table \ref{T12}  and  Table \ref{T14}, it is  found that all   $\chi^2$ coefficients are large. However,   almost all investigated FSD distributions become similar to the BL  distribution shown  in Table \ref{T14}. Thus,   the data set statistical quality becomes better after the deletion of abnormal value repetitions.

Nevertheless,exceptionally, the BL  $\chi^2$ test shows that the FSD distribution of financial (FIN) price index becomes less similar to the BL distribution  after adjustment. However, since the differences are  very small,   and  deliver  the lowest $\chi^2$ ($\sim 25$), the deletion of  the abnormal values of the financial price index  nevertheless leads to a more appealing  (or reliable)   financial data set  rationality. 

  \subsection{ Distance analysis} \label{distanceanal} 
Distance measures are   found to be smaller  when anomalous repetitions are removed.  
In addition, the high values of $d^*$ and $a^*$  show that the  adapted  distributions are  consistently more similar to the BL distribution.

Notice that  the $a^*$ of HEA seems  to be an outlier,  both before  ($\sim 0.28$) and after  ($\sim 0.22$) removing anomalous values. This is  in   contrast to the $a^*$ TECH value, which is an outlier  in the raw data  analysis, but turns out to be "very reasonable", after removing anomalous values.

  \section{Discussion } \label{subdiscussion} 
  
The present work has two aims. One is a test of Benford's law on specific data,  which addresses the question whether the aggregated data method  is reliable; 
 the second aim is  to touch upon the question whether Benford's law applies  to  aggregated data, in particular from emerging countries for various industry sectors.  An answer to the latter question is relevant before producing any research activity on comparing, for example, investments, growth, and financial indices of various sectors before economic policy or portfolio strategies.

It has been argued that it is of interest to analyze such financial data along BL lines. It is much agreed with e.g. Judge and Schechter   \cite{JudgeSchechter}   that BL  is  $definitively$ $not$  an indubitable  test of data exactness.  Thus, some  deviation from the Benford distribution would not provide a conclusive proof of " data manipulation", just as conformity does not prove cleanliness of the data.  Nevertheless, BL may  be considered useful as an aid in  analytical procedures of testing the exactness of  financial reports    
 \cite{ref[45]}.

Thus, if the test is  conclusive, analyzers  would be happy, but if not, this induces more questions and reflexions. The apparent  lack of agreement with BL for the 1-st digit of the raw data only, as found in the previous sections is somewhat frightening.  Fortunately, observing the existence of anomalous values leads to a more agreeable aspect of the data. Therefore one can conclude that the BL test  in Sect; \ref{sec:discussionBL} was useful on the original data.

 This leads to basically two sets of questions,  economic and financial ones, about the specificity of the aim, resulting from an accumulation of items: 
\begin{itemize} \item (i)  a first criticism should be on the aggregation method; one could demand more information on the items leading to the final sum of  values. One can indeed wonder about what was really  accounted for? Although the reported  values in each country pertained to a concluded year, and might concern different items, it might occur that   some rounding factor accumulated so much in a few cases as modifying the first digits of items and finally the global report.  Our argument is that we have examined such a possibility and took care as much as possible of a coherent scheme.
\item (ii) The anomalies might be only the result of sloppy, deliberate manipulation or unintentional but lazy accounting, quite in contrast from data manipulation by Governments 
 \cite{GER12.11.243}.  We have indeed considered such a possibility, and removed suspicious data.
\item (iii) A third hypothesis might  have a more fundamental aspect: indeed, non-conformity with Benford's law should not be qualified as a reliable sign of poor quality of macroeconomic data, but could rather be based on marked structural shifts in the data set   
 \cite{Gonzales-GarciaPastor09}. 
Gonzales-Garcia and Pastor  point out   that  (we quote)  {\it "rejection of Benford's law may be unrelated to the quality of statistics, .... . Hence, nonconformity with Benford's law should not be interpreted as a reliable indication of poor quality in macroeconomic data"}.   
 In some sense, this is a safe side confirming the need of a BL test before further research, and proposing a reflection on the data at the  first analytical  stage.
\end{itemize}


 \section{Conclusions}\label{sec:conclusions}
 
  We have looked at 10 industry sectors, in  6 top emerging countries, over a 15 year time span. The econophysics  analysis pertains to an accounting procedure along Benford's law   first significant digit  and to some statistical analysis of the distance between distributions.  It is shown that several  visually anomalous data  have to  be {\it a priori} removed.
 
 One  conclusion  is that complementary accounting techniques  tests should be considered before deciding whether some data is faked or erroneous.  
  Moreover, the results  indicate  that data reliability is a mandatory  aspect  to be observed  before proceeding  with further analysis and modeling. Some practical information has to be necessarily outlined before and after some  scientific analysis. 


\clearpage

 \newpage

 \begin{table}\begin{center} \begin{tabular}{|c|c|c|c|c|c| l |    }
\hline     
sector $s$	&	$N_s$	&	Mean	&	Std. Dev.	&	Min	&	Max	\\ \hline			
MATS	&	1080	&	0.288301	&	5.941751	&	-35.97069 	&	23.03240 	\\			
GDS	&	1007	&	0.480227	&	5.265518	&	-21.68574 	&	80.21888 	\\			
SVS	&	1080	&	0.242929	&	9.512246	&	-105.81600 	&	192.71100 	\\			
FIN	&	1080	&	0.364663	&	5.282102	&	-34.95486 	&	22.45599 	\\			
HEA	&	1060	&	0.270010	&	4.293489	&	-17.68409 	&	75.60660 	\\			
INDUS	&	1080	&	0.289500	&	5.331192	&	-34.29641 	&	48.92794 	\\			
OIL	&	963	&	0.189449	&	6.152058	&	-28.31262 	&	106.30770 	\\			
TECH	&	563	&	0.018170	&	5.288614	&	-25.36837 	&	45.13081 	\\			
TELE	&	1036	&	0.156854	&	5.138521	&	-35.17884 	&	62.35700 	\\			
UTIL	&	914	&	0.278857	&	4.494412	&	-28.60670 	&	26.05142 	\\ \hline			
 \end{tabular}  \end{center}
\caption{ 
Descriptive statistics of  raw FTSE aggregated data distributions according to industrial and financial sectors }  \label{T4}
\end{table}

  \begin{table} \begin{center} \begin{tabular}{|c|c|c|c|c|c| c|c|c|c|c|c|     }
\hline     
Time	&	MATS	&	GDS	&	SVS	&	FIN	&	HEA	&	INDUS	&	OIL	&	TECH	&	TELE	&	UTIL	\\ \hline	
19/06/00	&	393.85	&	306.97	&	350.87	&	352.27	&	353.19	&	326.67	&		&	324.88	&		&	346.67	\\	
19/07/00	&	419.54	&	318.72	&	375.72	&	390.77	&	392.12	&	332.86	&		&	301.05	&		&	372.63	\\	
19/08/00	&	672.12	&	357.51	&	424.42	&	504.35	&	441.34	&	300.67	&		&	342.92	&		&	475.36	\\	
19/09/00	&	551.57	&	323.42	&	316.22	&	359.97	&	{\bf 444.19}	&	232.00	&	345.30	&	315.77	&		&	404.89	\\	
19/10/00	&	436.36	&	263.26	&	291.00	&	333.02	&	{\bf 444.19}	&	217.52	&	316.80	&	236.61	&		&	387.36	\\	
19/11/00	&	514.74	&	318.45	&	349.38	&	440.82	&	{\bf 444.19}	&	254.57	&	374.48	&	323.73	&		&	468.56	\\	
19/12/00	&	483.85	&	301.04	&	341.78	&	469.70	&	{\bf 444.19}	&	253.05	&	330.55	&	309.58	&		&	472.67	\\	
19/01/01	&	494.17	&	276.96	&	365.06	&	494.72	&	{\bf 444.19}	&	253.83	&	351.20	&	307.70	&		&	490.06	\\	
19/02/01	&	571.19	&	293.50	&	392.49	&	476.44	&	{\bf 444.19}	&	256.90	&	423.46	&	277.65	&		&	560.58	\\	
19/03/01	&	641.28	&	496.10	&	410.26	&	661.23	&	{\bf 444.19}	&	395.62	&	436.04	&	337.65	&		&	701.20	\\ 	
19/04/01	&	796.57	&	638.30	&	464.53	&	934.58	&	665.35	&	508.53	&	465.78	&	425.80	&		&	792.83	\\ 	
19/05/01	&	837.03	&	686.18	&	512.03	&	989.45	&	643.00	&	561.10	&	543.03	&	425.03	&		&	849.41	\\ \hline	
 \end{tabular}  \end{center}
\caption{ 
An extracted part of the whole data set. This table covers relevant information only on China between June 2000 and May 2001, from Datastream. We provide this table as an example to visually demonstrate  abnormal repetitions  in the data set. They are  emphasized  with  bold font. The Table blank content also allows to point out that some data is  sometimes unavailable.} \label{T10}
\end{table}

     \begin{table}\begin{center} \begin{tabular}{|c|c|c|c|c|c| l |    }
\hline     
sector $s$	&	$N_s$	&	Mean	&	Std. Dev.	&	Min	&	Max	\\ \hline			
MATS	&	1080	&	0.288301 	&	5.941751 	&	-35.97069 	&	23.03240  	\\			
GDS	&	906	&	0.437259 	&	4.870018 	&	-21.68574 	&	24.15624 	\\			
SVS	&	992	&	-0.023972 	&	5.527875 	&	-62.08636  	&	19.45960  	\\			
FIN	&	1075	&	0.398876 	&	5.183369 	&	-26.34447 	&	22.45599 	\\			
HEA	&	724	&	0.266428 	&	4.324815 	&	-17.68409 	&	18.92318  	\\			
INDUS	&	1007	&	0.261899 	&	5.303363 	&	-34.29641  	&	21.00709  	\\			
OIL	&	816	&	0.051542 	&	5.376211 	&	-28.31262 	&	20.92057  	\\			
TECH	&	449	&	0.024091 	&	5.923321 	&	-25.36837  	&	45.13081  	\\			
TELE	&	1015	&	0.098663 	&	4.809565 	&	-35.17884  	&	22.98440  	\\			
UTIL	&	745	&	0.375603 	&	4.860502 	&	-19.12928  	&	26.051420 	\\ \hline					
  \end{tabular}  \end{center}
\caption{ 
Descriptive statistics of variables after the deletion of abnormal repetitious numbers.  The variables are the 10    industriesÕ aggregated monthly stock return  of the six countries mentioned in the text. As the table shows, the numbers of observations  $N_s$ have been decreased,  with respect to Table 1 data.
 } \label{T5}
\end{table}

     \begin{table}\begin{center} \begin{tabular}{|c|c|c|c|c|c| c|c|c|c|c|c|     }
\hline     
sector $s$	&	Obs	&	1	&	2	&	3	&	4	&	5	&	6	&	7	&	8	&	9	\\ \hline						
MATS 	&	1086	&	23.2044	&	18.3241	&	17.9558	&	13.2597	&	8.6556	&	5.5249	&	4.8803	&	4.4199	&	3.7753	\\						
GDS	&	1013	&	35.5380	&	14.1165	&	7.5025	&	8.1935	&	12.4383	&	4.7384	&	5.4294	&	6.1204	&	5.9230	\\						
SVS	&	1086	&	33.0571	&	14.5488	&	10.9576	&	7.7348	&	12.2468	&	5.5249	&	5.9853	&	6.9061	&	3.0387	\\						
FIN	&	1086	&	36.2799	&	24.1252	&	9.0239	&	4.6961	&	3.9595	&	5.8932	&	6.2615	&	4.9724	&	4.7882	\\						
HEA	&	1066	&	21.8574	&	4.2214	&	9.5685	&	9.0056	&	9.1932	&	7.4109	&	12.2889	&	10.9756	&	15.4784	\\						
INDUS	&	1086	&	20.9024	&	17.6796	&	22.5599	&	4.7882	&	5.2486	&	9.1160	&	10.4052	&	5.3407	&	3.9595	\\						
OIL	&	969	&	23.4262	&	26.4190	&	12.6935	&	8.5655	&	8.2559	&	4.8504	&	8.3591	&	3.7152	&	3.7152	\\						
TECH	&	568	&	41.7254	&	16.0211	&	11.9718	&	11.6197	&	9.1549	&	4.5775	&	1.7606	&	1.4085	&	1.7606	\\						
TELE 	&	1042	&	39.4434	&	15.6430	&	4.7985	&	5.5662	&	6.1420	&	4.9904	&	7.2937	&	8.6372	&	7.4856	\\						
UTIL	&	920	&	32.5000	&	9.8913	&	8.0435	&	19.0217	&	4.8913	&	9.0217	&	4.5652	&	6.4130	&	5.6522	\\ \hline						
FSD BL	&		&	30.1030	&	17.6091	&	12.4939	&	9.6910	&	7.9181	&	6.6947	&	5.7992	&	5.1153	&	4.5757	\\ \hline		\end{tabular}  \end{center}
\caption{ 
 First significant digit (FSD)   frequency of FTSE global price index for each industry  and  for each digit;  the  theoretically expected FSD value  from BL    is also given. } \label{T11}
\end{table}

     \begin{table} \begin{center} \begin{tabular}{|c|c|c|c|c|c| c|c|c|c|c|c|     }
\hline     
sector $s$	&	Obs	&	1	&	2	&	3	&	4	&	5	&	6	&	7	&	8	&	9	\\ \hline						
MATS 	&	1080	&	23.2044 	&	18.3241 	&	17.9558 	&	13.2597 	&	8.6556 	&	5.5249 	&	4.8803 	&	4.4199 	&	3.7753 	\\						
GDS	&	906	&	35.5380 	&	14.1165 	&	7.5025 	&	8.1935 	&	12.4383 	&	4.7384 	&	5.4294 	&	6.1204 	&	5.9230 	\\						
SVS	&	992	&	33.0571 	&	14.5488 	&	10.9576 	&	7.7348 	&	12.2468 	&	5.5249 	&	5.9853 	&	6.9061 	&	3.0387 	\\						
FIN	&	1075	&	36.2799 	&	24.1252 	&	9.0239 	&	4.6961 	&	3.9595 	&	5.8932 	&	6.2615 	&	4.9724 	&	4.7882 	\\						
HEA	&	724	&	21.8574 	&	4.2214 	&	9.5685 	&	9.0056 	&	9.1932 	&	7.4109 	&	12.2889 	&	10.9756 	&	15.4784 	\\						
INDUS	&	1007	&	20.9024 	&	17.6796 	&	22.5599 	&	4.7882 	&	5.2486 	&	9.1160 	&	10.4052 	&	5.3407 	&	3.9595 	\\						
OIL	&	816	&	23.4262 	&	26.4190 	&	12.6935 	&	8.5655 	&	8.2559 	&	4.8504 	&	8.3591 	&	3.7152 	&	3.7152 	\\						
TECH	&	449	&	41.7254 	&	16.0211 	&	11.9718 	&	11.6197 	&	9.1549 	&	4.5775 	&	1.7606 	&	1.4085 	&	1.7606 	\\						
TELE 	&	1015	&	39.4434 	&	15.6430 	&	4.7985 	&	5.5662 	&	6.1420 	&	4.9904 	&	7.2937 	&	8.6372 	&	7.4856 	\\						
UTIL	&	745	&	32.5000 	&	9.8913 	&	8.0435 	&	19.0217 	&	4.8913 	&	9.0217 	&	4.5652 	&	6.4130 	&	5.6522 	\\ \hline						
FSD BL	&		&	30.1030 	&	17.6091 	&	12.4939 	&	9.6910 	&	7.9181 	&	6.6947 	&	5.7992 	&	5.1153 	&	4.5757 	\\ \hline					\end{tabular}  \end{center}
\caption{ 
 First significant digit (FSD)  frequency  of FTSE global price index for each industry after the deletion of abnormal repetitious numbers; the  expected FSD  value according to BL is also given.  } \label{T13}
\end{table}

    \begin{table} \begin{center} \begin{tabular}{|c|c|c|c|c|c| c|c|c|c|c|c|     }
\hline     
sector $s$	&	Corr. coeff.	&	$\chi^2$	&	$M$	&	$d^*$	&	$a^*$	\\ \hline
MATS 	&	0.9092	&	40.5876**	&	0.06898	&	0.09054	&	0.00479	\\
GDS	&	0.9383	&	57.4226**		&0.05435	&	0.09129	&	0.01066	\\
SVS	&	0.9616	&	54.1127**		&0.04328	&	0.06584	&	0.00147	\\
FIN	&	0.9614	&	25.1805**		&0.06516	&	0.10787	&	0.05716	\\
HEA	&	0.4426	&	90.1797**		&0.13387	&	0.19834	&	0.27801	\\
INDUS	&	0.7543	&	39.9240**	&	0.10065	&	0.14564	&	0.05613	\\
OIL	&	0.8765	&	41.2437**		&0.08809	&	0.10870	&	0.00605	\\
TECH	&	0.9814	&	49.7493**	&	0.11622	&	0.12685	&	0.13523	\\
TELE 	&	0.9113 	&	39.6741**	&	0.09340	&	0.13053	&	0.02076	\\
UTIL	&	0.8469	&	36.5188**		&0.09330	&	0.12873	&	0.03812	\\ \hline
	\end{tabular}  \end{center}
\caption{ 
Pearson correlation coefficient,   $\chi^2$ test value, the $M$  distance and  other distance measures, as defined in the text, between BL and the investigated distributions; ** indicates a 99\% significant departure from BL. 
}\label{T12}
\end{table}   

    \begin{table}
    \begin{center} \begin{tabular}{|c|c|c|c|c|c| c|c|c|c|c|c|     }
\hline     
sector $s$	&	Corr. coeff.	&	$\chi^2$	&	$M$	&	$d^*$	&	$a*$	\\ \hline				MATS 	&	0.9092 	&	40.5876** 	&	6.8986 	&	0.0905 	&	0.0048 	\\			GDS	&	0.9640 	&	30.9417** 	&	5.6738 	&	0.0748 	&	0.0074 	\\				SVS	&	0.9723 	&	38.0961** 	&	4.2938 	&	0.0612 	&	0.0035 	\\				FIN	&	0.9599 	&	25.7035** 	&	6.6053 	&	0.1103 	&	0.0577 	\\				HEA	&	0.5483 	&	78.9243** 	&	12.9368 	&	0.1844 	&	0.2168 	\\				INDUS	&	0.8550 	&	35.4757** 	&	7.7164 	&	0.1128 	&	0.0657 	\\			OIL	&	0.9851 	&	40.2773** 	&	2.5545 	&	0.0395 	&	0.0068 	\\				TECH	&	0.9304 	&	56.5568** 	&	4.8145 	&	0.0865 	&	0.0598 	\\			TELE 	&	0.9170 	&	40.3456** 	&	10.1123 	&	0.1341 	&	0.0086 	\\			UTIL	&	0.9419 	&	31.9800** 	&	9.6048 	&	0.1125 	&	0.0135 	\\ \hline			\end{tabular}  \end{center}
\caption{ 
Pearson correlation coefficient,  $\chi^2$ test value, the $M$  distance and  other distance measures, as defined in the text, between BL and investigated distributions after data manipulation; ** indicates a 99\% significant departure from BL. 
 Comparing the distributions before and after adjustment through such coefficients shows that these distributions are more similar to the BL distribution, - except for  the financial (FIN)'s FSD distribution.} \label{T14}
\end{table}   
 
    \newpage


\newpage

\begin{thebibliography}{0}
\bibitem{Nigrini2012}  M. J. Nigrini, Benford's Law: Applications for forensic accounting, auditing, and fraud detections 
(John Wiley \& Sons, 2012).    

 \bibitem{ref[1]} S. Newcomb, Note on the frequency of use of the different digits in natural numbers, Amer. J. Math. 4 (1881) 39-40.

\bibitem{ref[2]} F. Benford, The law of anomalous numbers, Proc. Amer. Philos. Soc. 78 (1938) 551-572.

 \bibitem{PS8.11.1} A. Berger and T. P. Hill,  A basic theory of Benford's  Law, Probability Surveys  8 (2011) 1-126.

\bibitem{Mebane08}  W. R. Mebane Jr,   'Election forensics: the second-digit Benford's law test and recent American presidential elections.'  
 in {\it Election fraud: detecting and deterring electoral manipulation}, (Brookings Institution Press Washington, DC, 2008), pp. 162--81
 

  \bibitem{JudgeSchechter} G. Judge and L.  Schechter, Detecting Problems in Survey Data Using Benford's Law, J. Hum. Resour. 44 (2009) 1-24.
  
\bibitem{EPJB87.14.261MirRCMABenfordIT}   T. A. Mir,  M. Ausloos,  and  R. Cerqueti,  Benford's  law predicted digit distribution of aggregated income taxes: the surprising conformity of Italian cities and regions. Eur. Phys. J. B 87  (2014)   261.

\bibitem{Nigrinibook}  M. J. Nigrini, Digital analysis using Benford's law: tests statistics for auditors,  (Global Audit Publications, Vancouver, 2000)    

\bibitem{GER12.11.243} B. Rauch, M. G\" ottsche, G. Br\" ahler,  S.  Engel, Fact and Fiction in EU-Governmental Economic Data,  Germ. Econ. Rev.  12 (2011) 243-255. 

\bibitem{0607168befordlawbiblio} W. H\"urlimann, Benford's law from 1881 to 2006,  

$ http ://arxiv.org/abs/math.ST/0607168$  (2006)


 \bibitem{Kossovsky14}  A.E.  Kossovsky,    Benford's Law: Theory, the General Law of Relative Quantities, and Forensic Fraud Detection Applications. (World Scientific Publishing Co, 2014.)

 \bibitem{Kossovsky17}   A.E.    Kossovsky,  Small is Beautiful,   
  
  $https://www.amazon.com/Small-Beautiful-Numerous-Rare-World/dp/069291241X$   (2017).
  
    \bibitem{Ley96}  E. Ley, On the peculiar distribution of the U.S. stock indexes' digits.   Amer. Statist.  50 (1996) 311-313.
  
  \bibitem{DeCeusteretal98} M. J. K.  De Ceuster,  G. Dhaene,  and T. Schatteman,  On the hypothesis of psychological barriers in stock markets and Benford's Law,   Journal of Empirical Finance   5 (1998)    263--279 .

  \bibitem{KrakarZgela09}  Z. Krakar  and M.   \v{Z}gela,  Evaluation of Benford's Law Application in Stock Prices and Stock Turnover,    Informatologija  42  (2009) 158-165.

 \bibitem{Corazzaetal10}   M. Corazza, A. Ellero,  and A. Zorzi,    Checking financial markets via Benford's law: the S\&P 500 case, in     Mathematical and statistical methods for actuarial sciences and financ,  M. Corazza and C. Pizzi, Eds. (Springer-Verlag Italia,  2010) pp.  93--102. 
 
 \bibitem{ZhaoWu10} S. Zhao and W. Wu, 
 Does Chinese Stock Indices Agree with Benford's Law?   in Management and Service Science (MASS), 2010 International Conference on, pp. 1--3. 

  \bibitem{Zgela11}  M.    \v{Z}gela,   Application of Benford's law in analysis of DAX percentage changes, 
 Cybernetics and Information Technologies  11  (2011) 53-70.
  
    \bibitem{Karavardar14}  A. Karavardar,   Benford's Law and an Analysis in Istanbul Stock Exchange (BIST),  International Journal of Business and Management 9  (2014) 160-172.

  \bibitem{Cinko14} M. Cinko,   Testing distribution of BIST-100 returns by Benford Law,  Journal of Economics Finance and Accounting  1  (2014)  184--191. 

 \bibitem{Durtschietal04} C. Durtschi, W. Hillison, and C. Pacini,  The effective use of Benford's law to assist in detecting fraud in accounting data,  J. Forensic Acc. 5 (2004)  17-34.
  
\bibitem{CSF104.17.238MARCTAM/ATI_BL}  M. Ausloos, R. Cerqueti, and T. A. Mir, Data science for assessing possible tax income manipulation: The case of Italy, Chaos, Solitons and Fractals 104 (2017) 238-256.
    
 \bibitem{NigriniMiller09}  M. J. Nigrini and S. J. Miller, Data diagnostics using second-order tests of Benford's law. Auditing: A Journal of Practice \& Theory  28  (2009) 305-324. 
 
 \bibitem{ClippeAusloosTheil}
P. Clippe and M. Ausloos,  Benford's law and Theil transform of financial data. Physica A: Statistical Mechanics and its Applications 39 (2012) 6556-6567. 

 \bibitem{Gauvrit017AdvCognPsy} N. Gauvrit, J. Ch.  Houillon,  and J. P. Delahaye,  Generalized Benford's  Law as a Lie Detector. Advances in Cognitive Psychology 13    (2017) 121-127. 

 \bibitem{GunnelTodter09}   S.  Gunnel and K.  T\"odter,  Does Benford's Law Hold in Economic Research and Forecasting?, Empirica: Journal of Applied Economics and Economic Policy  36  (2009)  273-292.
 
 \bibitem{ref[45]} M.J. Nigrini and L.J. Mittermaier, The use of Benford's law as an aid in analytical procedures,  Auditing-J. Pract. Th. 16  (1997) 52-67.
   
     \bibitem{Nigrini96} M.J. Nigrini, A Taxpayer Compliance Application of Benford's Law, J.  Amer.  Tax Ass.  18 (1996)  72-91.
   
   \bibitem{MirAusJAIST}    T. A. Mir and M.  Ausloos,   Benford's law: a 'sleeping beauty' sleeping in the dirty pages of logarithmic tables, Journal of the Association for Information Science and Technology  (2017). doi: 10.1002/asi.23845;  http://arxiv.org/abs/1702.00554
  
   \bibitem{GrammetalPRE17}   R. Gramm,  J.  Yost,  Q.   Su,  and  R. Grobe, 'Applications of the first digit law to measure correlations', Phys. Rev. E 95   (2017)  042136 . 
  
 \bibitem{ChoGaines}  W.K.T. Cho and B.J. Gaines, Breaking the (Benford) Law: Statistical Fraud Detection in Campaign Finance', The American Statistician  61 (2007)  218-223.
     
      
 \bibitem{JM10}      J.   Mi{\'s}kiewicz,  'Entropy correlation distance method. The Euro introduction effect on the Consumer Price Index', Physica A: Statistical Mechanics and its Applications  389   (2010)  1677--1687.
   
   \bibitem{IMF2015}  International Monetary Fund  World Economic Outlook Database 2015  
 
  $https://www.imf.org/external/pubs/ft/weo/2015/01/weodata/index.aspx. $
   
    \bibitem{Gonzales-GarciaPastor09} J. Gonzales-Garcia and G. Pastor,  Benford's Law and Macroeconomic Data Quality, International Monetary Fund, Working Paper, 2009  
 
  {\it http://papers.ssrn.com/sol3/papers.cfm?abstract$_-$id=135643}
   


  



  
 \end{thebibliography}
\end{document}